\documentclass[authoryear,review,12pt]{elsarticle}

\usepackage{amssymb}

\usepackage{array}
\usepackage{ragged2e}
\newcolumntype{P}[1]{>{\RaggedRight\hspace{0pt}}p{#1}}

\journal{arXiv}

\begin{document}

\begin{frontmatter}



\title{Examining inverse generative social science to study targets of interest}


\author[inst1]{Thomas Chesney}

\affiliation[inst1]{organization={Nottingahm University Business School},
            addressline={Wollaton Road}, 
            city={Nottingham},
            postcode={NG8 1BB},
            country={UK}}

\author[inst2]{Asif Jaffer}
\author[inst1]{Robert Pasley}

\affiliation[inst2]{organization={Karachi Institute of Business Administration},
            addressline={University Road}, 
            city={Karachi City},
            postcode={Sindh 75270},
            country={Pakistan}\\,
            state={/*authors listed alphabetically}}

\begin{abstract}

We assess an emerging simulation research method---Inverse Generative Social Science (IGSS) \citep{Epstein23a}---that harnesses the power of evolution by natural selection to model and explain complex targets.

Drawing on a review of recent papers that use IGSS, and by applying it in two different studies of conflict, we here assess its potential both as a modelling approach and as formal theory.

We find that IGSS has potential for research in studies of organistions. IGSS offers two huge advantages over most other approaches to modelling. 1) IGSS has the potential to fit complex non-linear models to a target and 2) the models have the potential to be interpreted as social theory.

The paper presents IGSS to a new audience, illustrates how it can contribute, and provides software that can be used as a basis of an IGSS study.
\end{abstract}




\end{frontmatter}


\section{Introduction}
In response to \citep{Melnyk23} calling for a refresh of simulation studies organisation studies, we assess an emerging simulation research method---Inverse Generative Social Science (IGSS) \citep{Epstein23a}---that harnesses the power of evolution by natural selection to model and explain complex targets. IGSS has the ability to explore a vast search space of variables and their relationships seeking non-linear models to fit to data. More importantly the best models found have the potential to act as formal theory and as such bring with them the many advantages thereof, while also answering calls for more formal theory in studies of organisations \citep{Chesney21, Vancouver20}. Drawing on a review of recent papers that use IGSS, and by applying it in two different studies of conflict, we here assess its potential both as a modelling approach and as formal theory.

IGSS marries the power of evolutionary computing \citep{Holland92}---specifically here we use genetic programming \citep{Koza92}---and agent-based modelling \citep{Gilbert05}. Genetic programming is used to evolve micro-specifications---hereafter referred to as \emph{rules}---that when inserted into an agent model's code, dictate individual agents' behaviour. When such a model is run, if it generates macro behaviours that have been observed in a target of interest then under the paradigm of generative or `bottom up' social science \citep{Epstein96}, that target is considered to have been explained. We evaluate this view in Section \ref{formalTheory}. Before that, in Section \ref{overview} we outline IGSS. Then in Section \ref{conflictModels} we present two IGSS studies of conflict. A discussion of the work is offered in Section \ref{discussion}.

\section{Overview}\label{overview}
A typical IGSS study will progress as follows. There exists a target for which an explanation is sought. There also exist data on that target---or such data is collected---called the reference dataset. The reference dataset is input to software that evolves rules using genetic programming. The goal is to find rules that when implemented and run in an agent-based model will simulate the target such that it behaves as described in the reference dataset. This progression is unpacked over the proceeding paragraphs.

The target is a phenomenon or situation of interest that is being studied. The reference dataset stores observations on it capturing some interesting aspect of how it behaves. It is this behaviour that is to be explained by evolved micro rules. The reference dataset is quantitative and contains variables that have been identified as being theoretically relevant, or at least are suspected of such. This is the same as any quantitative research dataset. Indeed the reference dataset may have been collected previously for another study.

Agent-based modelling and genetic programming are well known in the organisations and decision support literature (see for example: \citep{Hajmohammad20, Shevchenko20, Braune22, Shady22, Chesney17, Zaffar19}). Agent-based modelling simulates a target allowing it to be experimented on, explored and observed \citep{Chesney21}. Genetic programming is the automatic evolution of computer code \citep{Banzhaf98}. To combine them requires genetic programming to evolve rules that will then be used to dictate agent behaviour in an agent model. (In addition the agent model will likely be required during evolution to test evolved rules for fitness.)

Genetic programming starts with a population of random (and therefore almost certainly meaningless) rules--the first generation of rules. For example:

\begin{enumerate}
\item IF agentAttribute1 $>$ agentAttribute2 THEN behaviour1
\item IF agentAttribute3 == globalVariable1 THEN behaviour2
\item IF agentAttribute1 $/$ agentAttribute2 $<$ agentAttribute3 THEN behaviour1
\item IF agentAttribute4 != globalVariable2 THEN behaviour3
\end{enumerate}

In this example Behaviours 1, 2 and 3 are each one possible action that an agent might take. Agent attributes are variables such as an agent's age or a summary of the situation they are currently in, perhaps the number of other agents that are close by. Global variables hold the same value for all agents. Examples could include a count of the number of agents, the current simulated weather, or the legal regulations under which the agents are operating.

Next each of the rules is tested for fitness. Fitness is the answer to the question: if agents follow this rule, how close will the output of the agent model be to the reference dataset? A common fitness metric is the squared difference between agent model data and reference data. It could be (and often will be) that to calculate fitness the agent model must be run which can make the process of completing an IGSS study take many weeks of computer processing. Three genetic operations are then used repeatedly to create the next generation of rules:

\begin{enumerate}
\item One of the rules (chosen with probability proportional to its fitness, so most likely a rule with high fitness) is allowed to reproduce and simply moves into the next generation.
\item One of the rules (again one most likely that has high fitness) mutates randomly and the mutated rule moves into the next generation. A mutation means that part of the rule is changed at random.
\item Two of the rules (as before two that most likely have high fitness) breed and the two child rules move into the next generation. This breeding is known as crossover which means that part of the first rule and part of the second rule are swapped with each other.
\end{enumerate}

These genetic operations allow for the essential ingredients of evolution to take place: reproduction, mutation and natural selection, which are known to be powerful tools for finding solutions \citep{Nowak06}. The second generation of rules is then tested for fitness. A third generation is created as the second was and so on until a stopping condition is reached.

So far few IGSS studies have been conducted but all have used the approach outlined above with small variation (the most notable of which is probably using genetic algorithms rather than genetic programming but the overall goal is the same). \citet{Greig23} investigate flocking behaviour in birds. Their reference dataset comes from a single run of an `ideal' agent-based model which was written based on observations of birds in flight \citep{Reynolds87}. \citet{Miranda23} studying field irrigation decisions by farmers use reference data collected from an experiment using human participants. \citet{Vu23} use genetic programming to match survey data in a study of alcohol consumption.

Creating a suitable test for fitness is often a challenging part of an AI effort but here IGSS has a huge advantage. Given that the goal is matching---not exceeding or extrapolating from or any other goal that might be assigned to an AI---the reference dataset, the necessary existence of a reference dataset means the fitness function can often simply be some measure of closeness, possibly Euclidean distance or as was suggested earlier, mean squared error.

This overview glosses over a lot of complexity--conducting an IGSS study is not a trivial matter. For example \citet{Vu23} test fitness against multiple objectives, not just one. The IGSS modeller must also develop the agent model that will use the evolved rules and decide on parameters such as the agent model's scale, what type of agents to include and the numbers of each, other initial parameter settings, and environment settings such as size and shape. The modeller may also need to make decisions about the discrete time nature of agent-modelling, specifically the point at which fitness is tested. This decision is very much context dependent. The bird flocking model of \citet{Greig23} for example is neverending: the agent birds are constantly flying with no end destination and therefore fitness can be tested at any point. This is not the case in the study by \citep{Vu23} of alcohol use and they test for fitness at a set point in time. 
Then after all of this, the final rule or rules that have been produced will need to be pruned and interpreted.

\section{Code as theory}\label{formalTheory}
We will use two phrases later that will be useful to define upfront. \emph{Language of the target} refers to descriptions of behaviour using terms that have meaning in the context of the target situation. If the target is salary negotiation, the language of the target might include `wages', `hardball', `play hard to get' and `hold your nerve'. \emph{Language of the target theory} refers to descriptions of behaviour using terms that have meaning in the theory being used to explain the target. This time for salary negotiations these might include `shared pot', `utility', `bounded choice', `rational' and so forth.

The word theory in social science is very imprecise, something that has long been criticised \citep{Sutton95}. Table \ref{terms} shows some of the ways it is used and the list is certainly incomplete. It is unlikely that this paper---or any other---will convince readers to agree on a meaning for theory, but we can make clear the definition that we use: theory is an explanation of data. Those data come from observing a target. Theory abstracts out elements of interest, as defined by the researcher, from a target and explains them. The theory will not attempt to explain more than these elements. (This is one reason why theory can be so poor when used for prediction.) Theory can appear in many formats. One format that will be relevant to our discussion later is computer code. (Again, perhaps not all readers will agree with this but a justification will be given at that time.)

The purpose of a social scientist is to contribute to theory--to explain. In most scientific endeavours, theory appears twice. When using an agent model as a research method, the first is that theory is used to develop a model, to dictate what micro behaviour is implemented. It is however rare that theory will be available to guide all micro behaviour. To make a simulation work, often additional behaviour---possibly even from additional theories---will have to be melded together. Discussing this in relation to using simulations for social science, \citet[p.98]{van08} say: \emph{The use of such terms as `theory', `framework', `model', and `paradigm' in psychology and the social sciences is as informal as the models themselves. One person`s conceptual model is another person`s theory or framework...In psychology and the social sciences, theorizing about a problem typically begins with verbal conceptual models, which then may be elaborated and adjusted over time as relevant empirical data accumulate. Formal mathematical models, computational models, statistical models, etc. rely on verbal conceptual models to specify variables and relations among them, although a host of extra assumptions and plausible estimates are typically needed to translate a verbal theory into a workable implementation.}

The exception to this is exploratory research where theory is not used, or not relied on as much, in building a model. Instead, when used in agent modelling, researchers are keen to explore different ideas they have for what might be going on in a target. These ideas do not have to be grounded firmly in theory and are implemented as micro behaviour to allow a researcher to observe whatever macro behaviour emerges from them.

The second time theory appears in scientific endeavours is as an output. The contribution a scientist makes is to theory--developing a new theory; adapting, confirming, or validating an existing theory. Theory is both an input and an output; theory goes into science, theory comes out of science.

Although rare in social science, it is possible that theory will be formal \citep{Adner09}. It is recognised that formalising theory brings many advantages \citep{Edwards10, Davis07, Vancouver16, Vancouver10, Farrell10, Busemeyer10, Lomi01, Grand16}. Formats for this include formal logic, mathematics, and computational models (which because they are used to simulate data ultimately means they are computer code). Examples of formal logic are difficult to find, see \citep{Kamps99}. Mathematics is of course common if we include results from linear regression and structured equation modelling, although \citep{Sutton95} for example would not consider this alone to be theory. Mathematical modelling work such as that used by for example \citep{Gupta20} is common although this method does blur with computational models and there is a question over whether writers who use this approach actually mean for their mathematical models to be interpreted as theory (though perhaps this is not an essential characteristic of theory). An entire field is devoted to computational models of social behaviour, for examples see the Journal of Artificial Societies and Social Simulation.
 
\citet{Vancouver20} demonstrate the process of turning imprecise theories into computational models and discuss why this is valuable. By implementing a theory as a computational model such as an agent-based model it can be tested and explored. Running a model generates data that can be compared with observations. This will demonstrate---prove in fact---that a model explains---or does not explain---those observations (it doesn't prove that it is the \emph{correct} explanation, only that it is \emph{an} explanation). Implementing a theory as a computational model makes an imprecise social science theory formal and precise, and allows for rigorous testing \citep{Edwards10, Edwards10a, Adner09}.

IGSS sits in the generative social science paradigm as described by \citet{Epstein96}. They say, ``\emph{we consider a given macrostructure to be `explained' by a given microspecification when the latter's generative sufficiency has been established}'' \citep[p.177]{Epstein96}. Under IGSS, the final chosen rule that has been selected from the evolved rules (which will have high fitness but might be the most fit--this is explained later in this section) should explain the target. If theory is an explanation, then this rules should at least have the potential to be considered theory--we would argue that it \emph{is} theory.

\citet{Vancouver20} would likely agree. Perhaps \citet{Sutton95} would too. Although they neither approve of or discount computer code as being theory, citing a lack of agreement about whether a model can constitute theory (p.371), agent model code does seem to fit well with their description of what strong theory is. Agent model code answers the question of why behaviours emerge. The code makes explicit the connections among elements in the target. The models present causal relationships and identify the timing of events. Code presents underlying processes that allow us to understand ``systematic reasons for a particular occurrence or nonoccurrence'' and present the microprocesses involved. A counter argument to this last point might be that agent-based models do not concern inter-relationships among macro level variables (although we think this is incorrect--the output from agent models do allow us to examined macro level variables). \citet[p.378]{Sutton95} finish their section with a quotation from Karl Weick: a good theory ``explains, predicts, and delights''. There is certainly something delightful about observing a phenomena emerge `in front of your eyes' on a computer screen while running an agent model.

At least in the sense of explaining social processes of interest, agent-based models can be considered as theory \citep{Chesney21} although as we say this we remember that not all readers will agree on what theory is. An agent model's code can be pruned to test and observe how robust the output is, thus potentially simplifying a theory. In addition, a model/theory can be explored to generate hypotheses that have not been thought of previously, and computational theories can be tested in ways that would be expensive, unethical, or impossible otherwise \citep{Chesney21, Vancouver20, Vancouver10}. As such agent-based models offer real possibilities for theory advancement that deserve more attention in this regard, see for example \citep{Oliveira21}.


We now consider how this idea of formal theory fits with IGSS. IGSS is best seen as exploratory research. Exploratory research is used to study a phenomenon that has not been studied previously. It is used to identify research questions for further examination and hypotheses that will later be tested. An exploratory researcher will likely get to prioritise these questions and lay out the future research agenda. As for statistics, exploratory cannot use null hypothesis significance testing. To do so would be a form of data dredging. (To explain why briefly: if 20 random ideas are tested using a significance threshold of 5\%, it would be expected that 1 of them would be observed because 5\% equates to a 1 in 20 chance. Therefore if 20 ideas that might be real effects are explored in an agent model, but in fact all of them are just random, one of them will probably wrongly be identified as being real.) Instead, exploratory research relies on graphical methods, descriptive statistics such as measures of central tendency, spread and correlation, and qualitative approaches. Many of these, such as interviews and focus groups, are inappropriate for agent models. Instead, when traditional agent modelling is used for exploratory research, it involves exploring ranges and combinations of parameter settings, and less frequently exploring a small number of competing rules.

The jurisdiction of IGSS is this later space--exploring competing sets of rules, many more than could possibly be explored manually. Indeed, IGSS can be considered `turbo charged' exploratory research. The consequence of this is that IGSS should firmly be considered exploratory research. Indeed if our interest was in implementing existing theory as an agent-based model then we would not have need for IGSS. Even if theory is guiding the design of the agents' environment, or guiding the selection of which agent breeds to include or the numbers of each to have, eventually the genetic programming algorithm must take over and learn micro rules on its own, at which point it will not be being guided by existing theory--it will be \emph{exploring} its search space. The consequence is that theory as an input is less relevant to IGSS than to other applications of agent models.

There are a infinite number of micro rules that will produce any macro behaviour perfectly and it would be impossible to decide between them simply by qualitatively or quantitatively observing the output of a model. When discussing this, \citet[p.45]{Chesney21} uses the analogy of rolling a die: \emph{Imagine you are holding a loaded die. If it is thrown on the ground, six comes up most frequently, followed by five, then four, and so on down to one. If however it is thrown against a certain slope, the frequency of getting a six is reduced. In fact if you throw the die now, each side comes up about 16 times in 100, exactly as expected from a normal die. Is the die now uniform? Or more simply, is it right? The system of the die's use has now produced a fair die even though we have not actually changed what most people would see as the `correct' part of the system that should have been changed.}

This issue is dealt with explicitly in extant IGSS studies. \citet{Greig23} recognise the problem thus: ``it is important to clarify that different solutions in program space can potentially map to the same solution in logical space'' and state that the solution will take a ``new set of non-trivial tools''. \citet{Miranda23} list their 10 best rules which are quite different from each other and there is no obvious way to choose between them other than closeness to the reference dataset, which as with the die metaphor does not necessarily give the `correct' solution. Interestingly they examine 10,000 rules using factor analysis to try to identify common elements (they refer to them as `strategies') found within them. This is a very exciting approach and is discussed further later.

\citet{Vu23} acknowledge the problem and deal with it in part by supplying their genetic algorithm with a ``library of theory building blocks'' which are ``entities, attributes and mechanisms''. These come from an existing computational model \citep{Buckley22} which is itself based on the Theory of Planned Behaviour \citep{Ajzen91}. As part of their discussion they state that the fact that the genetic algorithm uses all of the building blocks demonstrates that they ``are all of theoretical importance for determining individuals' intentions'' to behave in a certain way (in their case drinking alcohol). They then (Paragraph 6.4) go on to list additional empirical evidence as support of the building blocks' importance. This of course gives one of the ways in which the infinity of models is reduced, by giving the IGSS algorithm access to theoretically valid constructs to work with. The pity in this is that there may exist constructs that hadn't yet been considered by human researchers that the AI will not be given opportunity to find.

\citet{Epstein23a} and \citet{Epstein14} discuss the concept of cognitively plausible agents. This means that fitness is not enough to judge evolved rules. A rule has to be plausible. It must be possible to interpret it meaningfully in either the language of the target or the language of the target theory and in Section \ref{discussion} we discuss this in relation to our models from Section \ref{conflictModels}. \citet{Epstein23a} points out that a purely rationale agent for example would not be cognitively plausible and suggest that agent emotion, deliberation and social influence must contribute to interpreting and even guiding evolved behaviour.

Genetic programming is favoured for IGSS rather than many other forms of AI as it produces output that has strong potential to be meaningfully interpreted (although it is not the only possibility--a decision tree is another candidate). Contrast this with the output of an algorithm which trains a neural network which would be an interconnected series of weights something that could not possibly be interpreted meaningfully.

\begin{table}
\begin{center}
\begin{tabular}{P{5.5cm}P{4cm}P{2.5cm}} 
\hline
 Meaning & Example & Reference \\ 
\hline
 academic discipline & gender theory & \cite{Jule14} \\
 research method or paradigm of study & grounded theory & \cite{Martin86} \\
 topic of study and research method & game theory & \cite{Camerer11} \\
 a belief & marxism or marxist theory & \cite{Singer00} \\
 an umbrella term for a group of theories & theory of evolution & \cite{Nowak06} \\
 an explanation of data & this is the definition favoured here & \cite{Chesney21} \\
 answers to normative questions & theory of how society should be & \cite{Ougaard13} \\ 
\hline
\end{tabular}
\caption{Uses of the word `theory' in social science.}
\label{terms}
\end{center}
\end{table}

\subsection{Metatheory}
We end this section with a comment on a possibility that may be of interest to researchers. Rather than generating a theory, IGSS generates \emph{candidate} theories, competing alternative explanations. This suggests a role for IGSS in the study of metatheory. Metatheory is theory on theory, or theorising about the development and refinement of theories. A theory explains why certain behaviours are observed in a target. There exists an underlying process that creates those behaviours. It is this process that is being explained as it reveals itself in the target. With a traditional agent-based model, agents are told what this process is. With IGSS this behaviour is evolved in what was earlier called `turbo charged' exploratory research. This is valuable in and of itself but it will also allow for the study of how rules evolve; such a study would contribute to metatheory, a theory about the development of a theory, an explanation of where rules come from. We can picture IGSS as building a model of models to examine under what conditions a certain observation that is explained by a certain theory emerges.

For example, the factor analysis carried out by \citet{Miranda23} might be considered to be an early instance of this. Factor analysis is the authors' term and might be misleading. They use `factor' to refer to a set of commands that are applied to a number of parameters to produce a return value \citep{Gunaratne20}. In social science terms examples of factors would a construct such as personality, age, height, or a behaviour strategy such as playing tit-for-tat. From the 10,000 rules that were evolved, they assess how each factor and how joint factors (factor interaction) impact on model fitness by measuring joint contribution. This involves training a random forest regressor on the factor presence to fitness data and assessing this using tree analysis software \citep{Saabas19}. The most important factors then are the theoretical elements that when included lead to a theory being successful. We suggest that this can be considered the beginnings of a study of metatheory.

\section{Studies of conflict}\label{conflictModels}
Understanding and dealing with conflict is an important part of research into organisations. To illustrate IGSS and guide our discussion, in this section we 1) evolve behaviour in a hawk-dove game and 2) re-create the well known civil disobedience model.

\subsection{Hawk dove model}
For the hawk-dove agent model, in each time period agents move to a location upon which is a certain amount of a desirable resource. Locations can accommodate up to two agents at a time. Each agent must assess their situation and decide how much of the resource to try to take. If there is conflict between two agents over a resource then both get zero, otherwise they take what they wanted. One run of the simulation lasts 100 time periods. The decision of how much resource to try to take is the rule that is evolved using IGSS with the format:

IF condition THEN take x ELSE take y

Agents initially choose a location at random and will return to it unless they receive zero, in which case they choose another location at random. An agent's situation is recorded as a number of variables that are then used to construct the rule. These are:

\begin{enumerate}
\item amount of resource on my current location, $resource$
\item number of agents on my current location, $agents$
\item amount of resource I took in the previous round, $previousTook$ 
\item amount of resource on my location in the previous round, $previousResource$
\item number of agents on  my location in the previous round, $previousAgents$
\item number of agents in the simulation, $totalAgents$
\item total amount of resource I have taken, $totalResource$
\end{enumerate}

In this study we created reference datasets manually to capture various scenarios of interest. We looked at: equality (where all agents end up with the same total amount of resource, roughly the maximum possible for each) and several forms of inequality (using the wealth distribution in the UK described by The Equality Trust as a guide). The study was used to answer the question: how would all members of a society have to act in order to achieve this distribution of wealth? The code is written in R and is available at github.com/ThomasCNotts/RIGSS.

\subsubsection{Results}
To achieve near perfect equality, IGSS found the following rule:

\begin{itemize}
  \item[] IF (previousResource - previousResource) * (previousResource - previousResource) $>=$ (previousTook - resource) - (totalResource - agents) THEN take = 1
\end{itemize}

Evolved rules almost always have to be pruned to remove redundancy. In this case, the rule prunes down to `always take 1' which given the amount of resource and the number of agents in the simulation is a simple way to achieve equality. (The rule actually reduces to `if zero is greater than a negative number then take 1'.)

\begin{figure}
\includegraphics[scale=0.3]{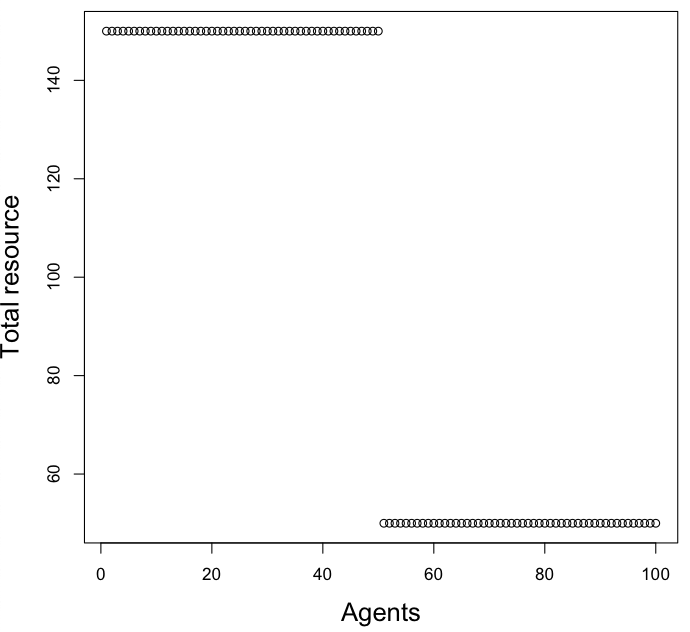}
\includegraphics[scale=0.3]{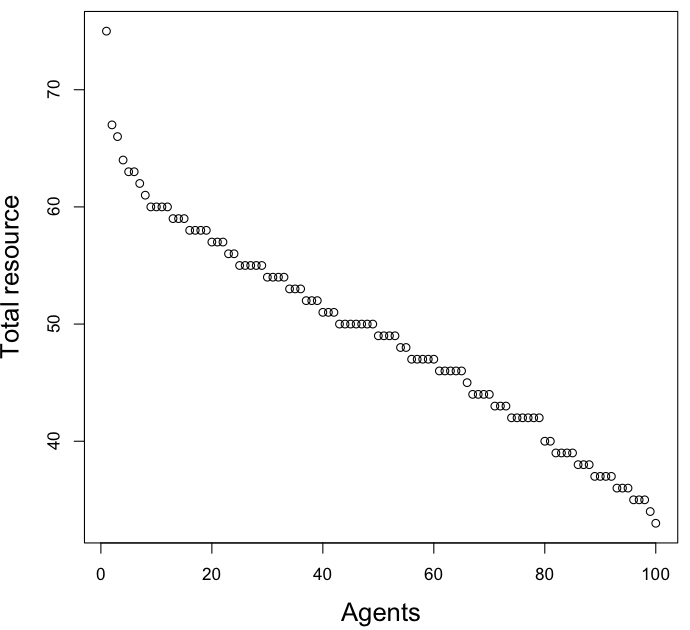}
\caption{The left panel shows wealth inequality with two extremes, split between two halves of society. The right panel shows the inequality produced by IGSS.}
\label{hawkdove}
\end{figure}

For inequality we started with a reference dataset that produces the distribution of wealth shown in the right panel of Figure \ref{hawkdove}. To try to match this IGSS found the rule:

\begin{itemize}
  \item[] IF (previousTook - totalAgents) != (agents AND previousTook) THEN take = 1 ELSE take = 9
\end{itemize}

which with pruning reduces to:

\begin{itemize}

  \item[]IF previousTook $>= 1$ THEN take = 1 ELSE take = 9
\end{itemize}

and produces the distribution shown in the left panel of Figure \ref{hawkdove}.




\begin{figure}
\includegraphics[scale=0.3]{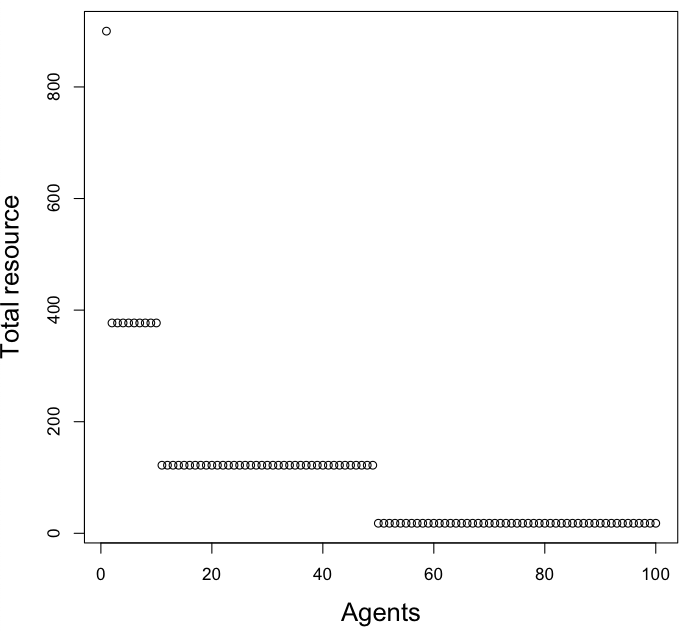}
\caption{A reference dataset that is closer to the real wealth inequality in the UK.}
\label{UKinequality}
\end{figure}

\subsection{Civil disobedience model}
The second illustrative model was written in NetLogo which is perhaps less well known than R. A specialist agent development environment, NetLogo is free to download. The civil disobedience agent model \citep{Epstein02} models violence flaring up in a densely populated environment. A version called Rebellion.nlogo is included in the NetLogo models library. Citizen agents are either happy or unhappy with their government, and are worried about arrest if they protest with violence. They calculate their political grievance based on an individual perceived hardship parameter and a government legitimacy global parameter, and calculate their chance of arrest based on the number of other people rebelling and the presence of police. If an agent’s grievance exceeds the risk of arrest by a small threshold, the agent decides to rebel. Occasionally, given certain initial parameters, there is a city-wide riot; but what other mirco rules would lead to the same behaviour? An IGSS study is a suitable way to explore this question.

There are three rules governing behaviour:

\begin{enumerate}
\item Rule M: if there are any new sites free, citizens move to a new site if they are not in jail; police move to a new site.
\item Rule A: citizens who are not in jail decide if they should riot. The rule for this is: riot if (grievance - risk-aversion * estimated-arrest-probability) $>$ threshold.
\item Rule C: if police sees a rioter or rioters, they arrest one of them randomly.
\end{enumerate}

The civil disobedience model has an advantage for researchers who want to run their first IGSS study in that the behaviour executed in each tick (the NetLogo term for a discrete unit of time) is independent of events that have happened in the previous tick--agents do not have to remember or factor in their past. This means that we can generate a static reference dataset and use this throughout learning. The alternative would be having to run the model every time we wanted to test the fitness of a generated rule as was the case with the hawk dove model which slows running considerably. To handle the A--M--C order of rules, we evolve each rule separately, one at a time.

We ran the model for 40 time periods using three different combinations of global variables, giving us data covering 120 time periods. For each time period we collected data on every agent, recording the agent variables. Since some internal agent states change more than once during a time period we captured certain agent variables at three different steps in each period. Essentially all data---including any hidden states---that determines behaviour was recorded in a csv file which became the reference dataset. A total of 122,160 data points were generated with 38 variables making up each point. These data were given as input to a second NetLogo program that implemented the genetic programming algorithm.

As before genetic programming was used to generate a candidate rule of the form IF condition THEN action. Due to bias in the reference dataset, in that in most time periods many more agents choose not to riot than choose to riot, we did not use a simple accuracy measure to test fitness. Instead we used a balanced accuracy measure as a fitness measure. This takes into account both sensitivity (true positive rate) and specificity (true negative rate) to provide a more comprehensive assessment of the model's ability to classify instances from different classes correctly.

Rule M, the move rule, is relatively simple and the algorithm
discovered a rule with perfect fit quickly. It is not qualitatively identical to the actual Rule M in the original Rebellion.nlogo but is quantitatively the same:

\begin{itemize}
  \item[]
IF 0 $>=$ jail-term0 AND threshold $<$ free-neighborhood0 [set movement-calculated 1]
\end{itemize}

Rule A is by far the most complicated of the three and it took much longer to achieve 98.07\% fitness:

Rule A:

\begin{itemize}
  \item[] IF ((movement-tracker0 = 0 AND movement-tracker1 = 1 OR active-binary $>$ 0) OR government-legitimacy * government-legitimacy $<$ perceived-hardship) AND (count cops-on neighborhood) * estimated-arrest-probability = jail-term [set active? True]
\end{itemize}

In the language of the target:

\begin{itemize}
  \item[] IF (((agent did not move last time AND agent moved this time) OR agent is active since last time) OR government-legitimacy squared $<$ hardship) AND count police * arrest probability = jail term of the agent THEN riot
\end{itemize}

Like Rule M, a version of Rule C was discovered with perfect fitness:

Rule C:

\begin{itemize}
  \item[]
if breed = cops [enforce]
\end{itemize}

\begin{figure}
\includegraphics[scale=0.3]{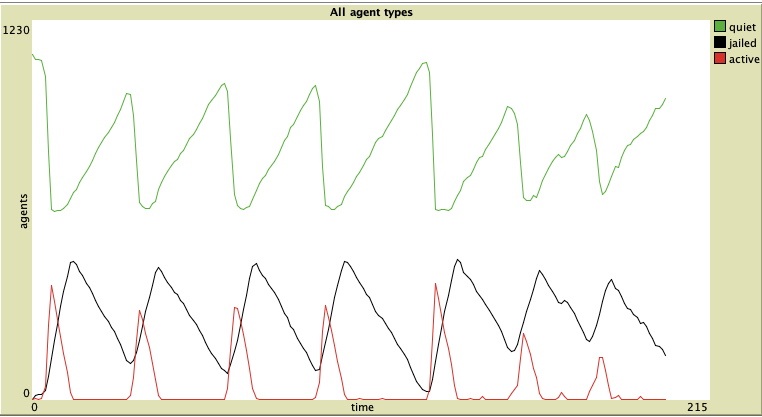}
\includegraphics[scale=0.3]{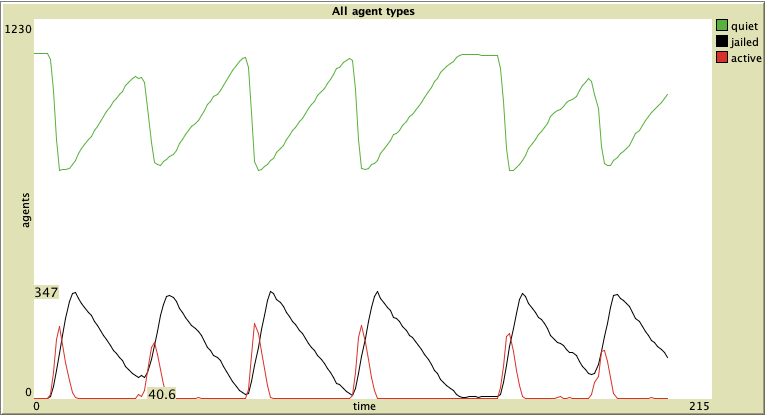}
\caption{The left panel shows the output from the original Rebellion.nlogo; the right panel shows the same output from Rebellion featuring our evolved rules.}
\label{compare}
\end{figure}

We created a new model with the new evolved rules and compared the results with the original model. The results are shown in Figure \ref{compare}. Qualitatively the two would be difficult to distinguish.

\section{Discussion}\label{discussion}
In 2009, \citet{Frigg09} published an examination of claims made about the philosophy of computer simulation arguing that far from demanding a new `metaphysics, epistemology, semantics and methodology', simulation raises few if any new philosophical problems. As IGSS is exploratory research it shares a philosophy of science with exploratory research. The fact that IGSS generates candidate theories quantitatively at a larger scale than `manual' exploratory research does not in itself bring fresh philosophical challenges. However there is much to get excited about. IGSS offers two huge advantages over most other approaches to modelling. 1) IGSS has the potential to fit complex non-linear models to a target and 2) the models have the potential to be interpreted as social theory.





One of the most interesting prospects is that the candidate theories IGSS generates can be \emph{non-linear}. Not only that but IGSS allows for the possibility of a candidate theory to be discovered made up of multiple non-linear behaviours belonging to different types of agents thus introducing heterogeneity between different types of agents. This can be compared with model discovery in the physical sciences where AI has long been used to search for linear and non-linear ordinary differential equations that yield an observable relationship between inputs and outputs (see for example \citet{Schmidt09}). The downside of this is that the search space will be enormous. While the review in this paper has highlighted several approaches to reducing it, this remains a avenue for future research.

Our illustrative examples raise many points to note about IGSS. Starting with specific results before moving on to make general comments about IGSS, for the hawk-dove model, while our system did not find a good fit for either of our inequality models, it did find a rule that leads to a smoother (and indeed therefore more realistic) unequal distribution of wealth (right side panel, Figure \ref{hawkdove}). The rule that gave allowed this to emerge exhibited extremes; in the language of the target it is that agents take either very little or a lot:

\begin{itemize}
  \item[] 
IF previousTook $>= 1$ AND agents $>= 1$ THEN take = 1 ELSE take = 9
\end{itemize}

According to the preceding argument, this rule can explain wealth inequality. This should be explored further in a fresh agent model to probe the relationships between extremes and wealth distribution. (Strictly as this would still be exploratory research, it would not have to be a fresh agent model although a fresh model would be needed if hypothesis testing were going to be used--see \citet[p.104]{Chesney21}. Indeed a brief exploration of this rule in our agent model reveals that any values where the second is greater than the first will lead to inequality, but the values above maximise it.) 


Looking at the civil disobedience model, Rule A as presented in the previous section may seem like it it is not explainable and this is a good case where the rule must be interpreted in the language of the target or the target theory. To recap the evolved version of the rule is:

\begin{itemize}
  \item[] IF ((movement-tracker0 = 0 AND movement-tracker1 = 1 OR active-binary $>$ 0) OR government-legitimacy * government-legitimacy $<$ perceived-hardship) AND (count cops-on neighborhood) * estimated-arrest-probability = jail-term [set active? True]
\end{itemize}

This can be interpreted in the language of the target as:

\begin{itemize}
  \item[] IF (((agent did not move last time AND agent moved this time) OR agent is active since last time) OR government-legitimacy squared $<$ hardship) AND count police * arrest probability = jail term of the agent THEN riot
\end{itemize}

These parameters can be further interpreted. For example, arrest probability = jail term is effectively arrest probability = 0. This rule then starts to make more sense. Human interpretation like this will be essential to make sense of complicated rules and as we have said before, this sort of interpretation is what a social scientists does so we should not fear it.


More generally, from our studies it is apparent that when running an IGSS study, there are a number of decisions that must be made almost arbitrarily. Each of these decisions could be a subject for further research. For example, the choice of primitives to give to the genetic programming algorithm to work with. Primitives are the mathematical operators from which the rules are built and there is very little theoretical guidance on what these should be. Other examples are the size of the initial population of rules and the probabilities of each genetic operation (that is, the chance of reproduction, mutation or crossover happening to a rule). In our hawk-dove model, the number of generations was chosen arbitrarily. Each of these decisions could potentially impact on results and yet there currently exists no systematic way to decide on all of them other than `rules of thumb'. We ran our hawk dove model for 100 time periods, the civil disobedience for 40--again these are arbitrary decisions that we suspect will have little bearing on results but this should be studied further.

Lastly we give some suggestions for possible IGSS studies in the area of decision support and organisations. IGSS could be used to model individual-level decision-making in supply chains (e.g., purchasing behavior, supplier relationship management) and simulate how these micro-behaviors aggregate to produce macro-level outcomes like market fluctuations, supply chain disruptions, or resilience. This could help predict and understand complex phenomena that emerge from individual interactions within the supply chain. IGSS could explore novel and potentially more efficient or ethical supply chain configurations that wouldn't be easily discovered through traditional optimization approaches. Finally IGSS could model how individual agents adapt their behavior when faced with changes like new regulations, economic crises, or technological advancements providing insight into potential unintended consequences of policies or predict how supply chains react to disruptions, aiding in better preparedness and policy design.

We finish with a final observation. An agent model is not a necessary part of an IGSS study. A reference dataset can be generated in any number of ways, only one of which is an agent model. A genetic programming algorithm could be run on data produced in an experiment, a survey, or an existing dataset to produce rules. It may be natural though not essential to then implement those rules in an agent model to demonstrate that they lead to the behaviour observed in the target; this is of course the idea behind generative social science. The main point is, that when IGSS software becomes easier to use, there will be many existing datasets that have be used to create many accepted theories that could be explored with IGSS to produce alternative theories that can then be assessed by researchers.

\bibliographystyle{elsarticle-harv} 

\end{document}